\documentstyle[secdot,12pt,epsf]{article}

\newenvironment{hangref}
  {\begin{list}{}{\setlength{\itemsep}{4pt}
  \setlength{\parsep}{0pt}\setlength{\leftmargin}{+\parindent}
  \setlength{\itemindent}{-\parindent}}}{\end{list}}

\begin{document}

\begin{center}

{\LARGE A Variable Depth Sequential Search Heuristic for the Quadratic
  Assignment Problem }\\[12pt]

\footnotesize

\mbox{\large Gerald Paul}\\
Center for Polymer Studies and Department of Physics,
590 Commonwealth Ave., Boston University, Boston, Massachusetts,
02215, USA,
\{\mbox{gerryp@bu.edu}\}\\[6pt]

\normalsize

\end{center}



\noindent We develop a variable depth search heuristic for the
quadratic assignment problem.  The heuristic is based on sequential
changes in assignments analogous to the Lin-Kernighan sequential edge
moves for the traveling salesman problem.  We treat unstructured
problem instances of sizes 60 to 400.  When the heuristic is used in
conjunction with robust tabu search, we measure performance
improvements of up to a factor of 15 compared to the use of robust
tabu alone.  The performance improvement increases as the problem size
increases.

\bigskip

\noindent {\it Key words:} combinatorics; quadratic assignment problem; 
hybrid heuristic; variable depth sequential search

\noindent {\it History:} 

\noindent\hrulefill
           







\section{Introduction}

The quadratic assignment problem (QAP) is a combinatorial optimization
problem first introduced by Koopmans and Beckman (1957).  It
is NP-hard and is considered to be one of the most difficult problems
to be be solved optimally.  The problem was defined in the following
context: A set of $N$ facilities are to be located at $N$ locations.  The
distance between locations $i$ and $j$ is $D(i,j)$ and the quantity of
materials which flow between locations $i$ and $j$ is $F(i,j)$.  The
problem is to assign to each location a single facility so as to minimize
the cost
\begin{equation}C=\sum_{i=1}^N \sum_{j=1}^N  F(i,j) D(p_i,p_j).
\end{equation}
where $u=p_i$ represents the assignment of facility $u$ to location
$i$.  We will consider symmetric instances of the problem ($
D(i,j)=D(j,i), F(i,j)=F(j,i), D(i,i)=F(i,i)=0$).

There is an extensive literature which addresses the QAP and is
reviewed in Pardalos et al. (1994), Cela (1998), Anstreicher (2003),
and James et al. (2009).  With the exception of specially constructed
cases, optimal algorithms have solved only relatively small instances
{$N \le 36$}.  Various heuristic approaches have been developed and
applied to problems typically of size $N\approx 100$ or less.

The most successful heuristics to date for large instances are {\it
  hybrid heuristics} that combine the {\it robust tabu search}, RTS,
(Taillard (1991)) with other techniques.  Here we propose a {\it variable
depth search} heuristic to be used in conjunction with RTS.  A  variable depth search is a local
search in which the number of moves along a search path is not fixed
but rather is determined by the likelihood that the search path will
be successful in finding an improved solution.  If a search path is
not promising, the search path is terminated and another path is
explored. Our {\it variable depth sequential search}, VDSS, is
inspired by the Lin-Kernighan variable depth search for the traveling
salesman problem (TSP) (Lin and Kernighan (1973) and Helsgaun(2000))

\section {Approach}
\label{approach}

It is helpful for this description to think of the
$N$ facilities and the matrix of flows between them in graph theoretic
terms as a graph of $N$ nodes and weighted edges, respectively.  Let us
denote
\begin{itemize}
\item{ nodes moved as $u_m, m=0,1,2,\ldots $}
\item{ location to which node $u_m$ is moved as $k_m.$}
\item{ location from which node $u_m$ is moved as $\ell_m.$}
\end{itemize}
The variable depth sequential search algorithm VDSS consists of a sequence
of moves of nodes $u_m$ from locations $\ell_m$ to locations $k_m$,
respectively.  The moves are sequential in that node $u_{m+1}$ is the
node located at location $\ell_m$. With each node as a starting node, the
moves are performed as a depth first search to a predefined depth,
$d$. In a given sequence of moves, we do not allow a node to be moved
more than once (see Appendix).

The sequence of moves terminates when either
\begin{itemize} 
\item {the sequence can be closed with move $u_z$ which results in
    an improved feasible solution by moving the node $u_z$ to
    the initial location of node $u_0$ or }
\item {a predefined limit on the number of moves that have been
    performed from a given start node is exceeded}.
\end{itemize}

Without any pruning of the search tree, the complexity of the search
would increase as $N^d$ where d is the depth of the search.  Therefor, we
prune the search tree as follows: With each move, we associate an
incremental gain $g_m$, the reduction in cost if the move is made.  We 
also define the cumulative gain $G_m$, the total gain after $m$ moves
\begin{equation}
G_m= g_1 + g_2 + g_3 + \ldots +g_m.
\end{equation}
Before a move is performed, the incremental gain is calculated and
added to the cumulative gain.  If the resulting cumulative gain will
be positive, the move is made; if not, the move is not made.

This pruning follows the pruning method introduced by Lin and
Kernighan (1973) for the traveling salesman problem (TSP) in which
edges, instead of nodes, are rewired sequentially.  They noted that,
for the TSP, if there is a series of sequential edge rewirings which
result in an improved solution there is a cyclic permutation of the
order of the rewirings for which the cumulative gain is always
positive.  Thus only sequences in which the cumulative gain after each move is
positive need be considered.  The Lin-Kernighan insight holds for the
TSP because the incremental gains from the rewirings are independent.
For the QAP, however, the insight may not always hold because the
incremental gains for the moves are not independent.  As a result, the pruning
may cause us to
\begin{itemize} 
\item{miss certain improved solutions which would be found if the pruning were not done or}
\item{proceed in the search but ultimately  not find a feasible improvement.}
\end{itemize}
Despite these issues, however, we find that the pruning approach
successfully reduces the complexity of the search while guiding the
search to improved solutions.

\section{Efficient Calculation of Incremental Gain}
\label{efficient}

Let $\Delta_0(u,k)$ be the incremental gain of a move of node $u$ from location $l$
to location $k$ before any moves in a sequence have been performed
\begin{equation}
\Delta_0(u,k)=2 \sum_{v=1,\ne u}^{N}[D(l_v,\ell)-D(\ell_v,k)]F(u,v)
\label{Delta0}
\end{equation}
where $\ell_v$ is the location of node $v$.  We can make the heuristic more
efficient by not having to calculate incremental gains using 
(\ref{Delta0}) each time they are needed.  This was done by Taillard
(1991) for the case of simple swaps.  We now describe how this is done
for sequential moves.

Because the incremental gains are not independent, $\Delta_0(u,k)$
will not necessarily be the incremental gain of moving $u$ to location
$k$ if other moves have been performed previously in the sequence.  If
other moves have been made, we can calculate, in constant time, the incremental gain of
the $n^{th}$ move in a sequence using
\begin{equation}
  g_n(u_n,k_n)= \Delta_0(u_n,k_n) + 2\sum_{m=1}^{n-1} (\delta_m^{\rm curr}-\delta_m^{\rm orig})
\label{Deltan}
\end{equation}
where
\begin{eqnarray}
\delta_m^{\rm curr}=[D(k_m,\ell_n)-D(k_m,k_n)]F(u_n,u_m)  \\
\delta_m^{\rm orig}=[D(\ell_m,\ell_n)-D(\ell_m,k_n)]F(u_n,u_m).
\end{eqnarray}
Here $\delta_m^{\rm curr}$ and $\delta_m^{\rm orig}$ are the contributions to
the incremental gain for the edge $(u_m,u_n)$ given the current and
original locations, respectively, of the node $u_m$.  Equation
(\ref{Deltan}) takes into account the fact that because of the previous $n-1$
moves the ends of the edges are no longer at the same locations as
they were before the sequence was started.  Using (\ref{Deltan})
instead of (\ref{Delta0}) reduces the VDSS time by a factor of $N$.

After a sequence of $n$ moves which results in a feasible improvement,
$\Delta_0$ is updated as follows: Assume the location of node $v$ is
$i$.  Then
\begin{eqnarray*}
 \Delta_0^{new}(v,j)=\Delta_0(v,j)
+ 2 \sum_{m=1}^n[(D(k_m,i)-D(k_m,j))\\
-(D(\ell_m,i)-D(\ell_m,j))]F(u_m,v),
\end{eqnarray*}
if $v$ is not one of the nodes moved in the sequence.  If $v$ is one
of the nodes moved, then $\Delta_0(v,j)$ must be calculated from
(\ref{Delta0}).  As with swaps (Taillard (1991)), the overall
evaluation of $\Delta_0$ takes $O(N^2)$ operations.  Because there are
relatively few sequences which result in feasible improvements, the
time spent in this calculation is a very small fraction of the total
processing time.

\section{Application of VDSS}

Sections \ref{approach} and \ref{efficient} describe the basic VDSS algorithm.  In
practice, in order to minimize the number of operations to find an
improvement, we execute VDSS with increasing maximum
depths ($d_1, d_2$, \ldots).  When an improvement is found at one depth, we start the VDSS
algorithm with the new optimal configuration at the smallest
proscribed depth.  If an improvement is not found at a given depth,
the algorithm is then executed at the next specified depth.  When a
pass is made through all depths with no improvement found, the run terminates.  
The code which implements VDSS is available in the Online
Supplement.

\section{Parameter Settings}

For RTS we use Taillard's code (available at
http://mistic.heig-vd.ch/taillard/codes.dir/tabou\_qap.cpp) with the
settings as described in Taillard (1991): number of iterations=$N^2$,
tabu list size between $0.9 N$ and $1.1 N$ and aspiration function
parameter=$2N^2$.  We run VDSS at two maximum depths: $d_1=2$ and $d_2=5$.
For any given start node, we limit the number of total moves attempts
to $10^5$.  These are the only tunable parameters.

\section{Computational Results}

We study the QAP instances Tai60a, Tai80a, and Tai100a from QAPLIB
(see Burkard et al. (1997)).  These instances are the most commonly
used recently for computational testing (James et al. (2009). The
instances are symmetric and the entries of the distance and flow
matrices are randomly and uniformly generated integers between 0 and
99 (Taillard  (1991)).  Additionally, to test our heuristic on larger
instances we create similar random instances of size $N=200$ and
$N=400$. Files containing these instances and the best known solutions
are included in the Online Supplement.

To compare the efficiency of heuristics, we use a time-to-target plot
(Aiex (2002), Aiex et al. (2002), and Oliveira et al. (2002)).  For
any given target solution value and the time to obtain that value, the
time-to-target plot plots the probability that the target cost will be
obtained. As described in Oliveira (2002), a solution target value is
first set.  The running time of the algorithm to achieve that cost or
lower is recorded.  This is done multiple times and the recorded times
are then sorted.  With the $i^{th}$ shortest time we associate a
probability $P_i=(i-1/2)/m$, where $m$ is the number of times recorded.

\begin{figure}[tbh]
\centerline{
\epsfxsize=10.0cm
\epsfclipon
\epsfbox{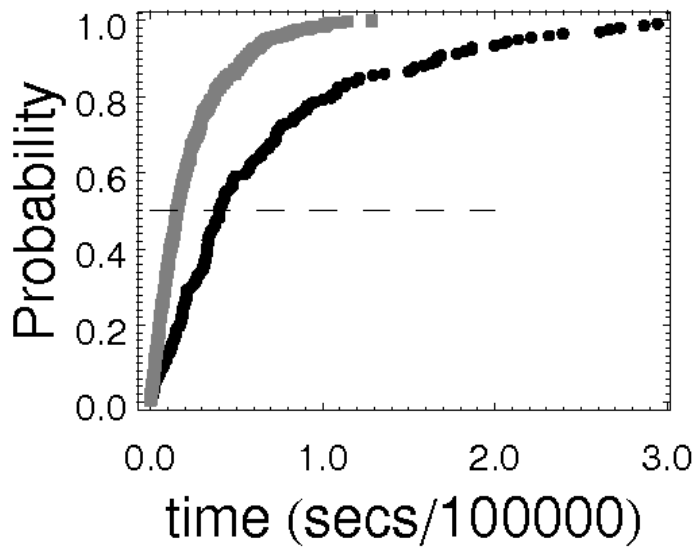}
}
\caption{Probability of obtaining target solution value of 21220000
  versus time for the Tai100a instance.  Lower (black) plot  is for
  RTS only runs; upper (gray) plot is for hybrid heuristic of
  RTS and VDSS.  The intersection of each plot with the dashed line at
  probability $0.5$ is the time $t_{50}$ used for the performance improvement factor calculation}.
\label{ttt}
\end{figure}

Figure \ref{ttt} compares time-to-target plots for RTS runs and hybrid 
heuristic runs composed of RTS followed by VDSS.  A single
hybrid heuristic run consists of first running RTS with a random
starting configuration and then running VDSS with the RTS solution as
input.  The instance tested is Tai100a and the target value is 21200000.  
We will see below that VDSS used alone is inferior to RTS.  However, as 
Figure \ref{ttt}illustrates the hybrid solution
of RTS + VDSS provides better performance than RTS alone.  

Let the time needed to achieve the target value with $50\%$
probability for a heuristic be $t_{50}$.  Then we define the
{\it performance improvement factor} of the hybrid heuristic (RTS + VDSS)
to RTS alone as:
\begin{equation}
I=\frac{t_{50}(RTS)}{t_{50}(RTS+VDSS)}.
\end{equation}
The performance improvement factor for Figure 1 is 2.61.  

\begin{figure}[tbh]
\centerline{
\epsfxsize=10.0cm
\epsfclipon
\epsfbox{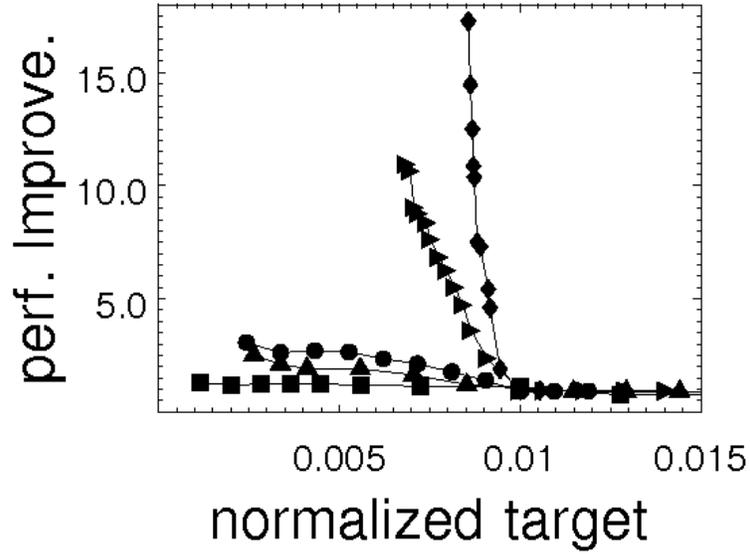}
}
\caption{Performance improvement versus normalized target values (see 
  (\ref{normalizedTarget}) ) for $N=60$ (squares), $80$ (up-triangles),
  $100$ (disks), $200$ (right-triangles) and $400$ (diamonds). Below the normalized target
  value of 0.01, the larger the value of $N$ the larger the slopes of
  the plots.}
\label{speedup}
\end{figure}

\begin{figure}[tbh]
\centerline{
\epsfxsize=10.0cm
\epsfclipon
\epsfbox{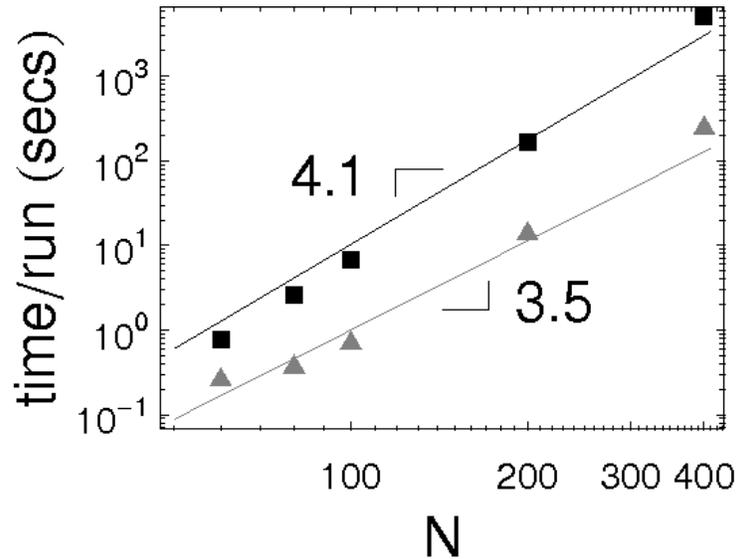}
}
\caption{Time per run versus $N$ for RTS (black upper plot)
  and VDSS (gray lower plot).  The slopes on these log-log
  plots are the power of $N$ of the computational complexity.}
\label{complexity}
\end{figure}

In Figure. \ref{speedup}  we plot the performance improvement factor for the 
instances studied versus various target values. Note that for each of
the plots, there is a target value above which the plots are essentially
flat; the performance improvement is constant and less than $1$ here because
RTS alone can reach these targets.  Below these {\it improvement thresholds} the performance improvement increases with decreasing target
values. In order to show the plots in a single figure, we plot performance improvement versus
normalized target values
\begin{equation}
\tilde \tau=\frac{\tau-b(N)}{b(N)}
\label{normalizedTarget}
\end{equation}
where $\tau$ is the non-normalized target value and $b(N)$ is a value
which results in the improvement thresholds of the plots being coincident.
Having the improvement thresholds coincident makes comparison of the plots
easier. We note that
\begin{itemize}
\item{for a given instance, the performance improvement factor
    increases as the target value is decreased because, as the 
    target is decreased, RTS alone is less and less likely to reach the 
    target. Running VDSS  following RTS, while adding time to a run,
    makes reaching the target more likely.}

\item{as the instance size increases, the performance improvement
    increases. Figure \ref{complexity} plots the time per run as a
    function of $N$ for RTS, and for VDSS when run following RTS. Note that for RTS the time
    scales as $N^x$ where $x\approx 4.1$.  This is consistent with the
    theoretical estimate that the complexity of RTS is $O(N^4)$ given
    that RTS requires $O(N^2)$ operations per iteration and $N^2$
    iterations per run.  For VDSS the time scales as $N^y$ where
    $y\approx 3.5$. This lower order complexity is one reason the
    performance improvement of the hybrid algorithm improves as $N$
    increases.  }
\end{itemize}

In Table \ref{table1} we list the instances tested and the results.
The lowest target value for which we determine the performance
improvement is determined by our ability to perform enough runs which
reach that target value to have meaningful statistics. Below the
target values shown in Table \ref{table1}, the time to perform the
needed number of runs would have been unreasonably long.  Performance 
improvement results shown in Table \ref{table1} are for these lowest target
values.

\begin{figure}[tbh]
\centerline{
\epsfxsize=10.0cm
\epsfclipon
\epsfbox{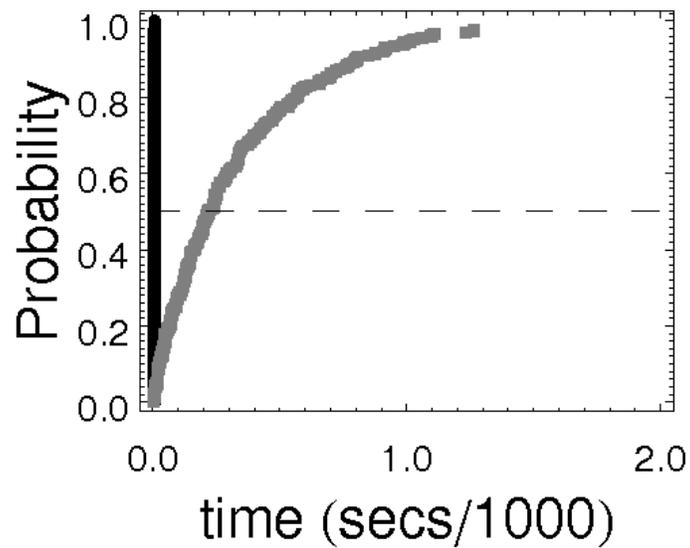}
}
\caption{Probability of obtaining target solution value of 21400000
  versus time for the Tai100a instance.  Black plot is for RTS-only runs; gray plot is for VDSS-only runs. The RTS plot appears essentially straight but would have a similar shape to the VDSS plot if magnified. 
 The intersection of each plot with the dashed line at
  probability $0.5$ is the time $t_{50}$ used for the performance improvement factor calculation}.

\label{tttlk}
\end{figure}

In Figure \ref{tttlk} we compare the performance of RTS alone and VDSS
alone.  As opposed to the hybrid heuristic RTS+VDSS, VDSS alone has
performance significantly inferior to RTS.

\section{Discussion/Conclusions}

We use the Lin-Kernighan pruning approach to implement an effective
variable depth sequential search which, when used in conjunction with
RTS, provides considerable performance improvement over
RTS alone.  RTS efficiently finds a local minima and VDSS explores the
neighborhood around this local minima to improve the solution.  Because RTS 
is a basic building block in a number of hybrid
heuristics, the combined RTS+VDSS approach described here may provide further performance improvements for those heuristics which use RTS alone.

To the best of our knowledge this is the first use of the Lin-Kernighan
pruning technique for a problem in which the incremental gains for the
sequential moves are not independent.  Our results raise the question
of whether variable depth sequential search with Lin-Kernighan pruning can be applied to other
problems in which incremental gains are also not independent and for
that reason the Lin-Kernighan approach may have not been applied to
them.  The code which implements VDSS is available in the Online
Supplement and can be used as a model for application to such other problems.

In addition, we introduce two new QAP instances, Pau200a and Pau400a,
available in the Online Supplement.  As processing speed increases, we expect 
that there will be a need for instances of this size for QAP heuristics.

\section*{Appendix -  Moving a node more than once in a sequence}

If a node is not allowed to be moved more than once in a sequence,
there are certain assignments of nodes to locations which cannot be
obtained by a single series of sequential moves starting with a given assignment.  For
example, the permutation 2,3,1,5,6,4 cannot be transformed into
1,2,3,4,5,6.  The transformation can be obtained with two separate
series of sequential moves each without moving a node more than once, but
if neither of these series of moves results in a positive gain, the
pruning algorithm will not allow these moves to be made.  If moving a
node more than once is allowed, any assignment can be obtained.  In
practice we have found that there is minimal benefit, if any, to
allowing nodes to be moved more than once and have not allowed it in
our runs.

\section*{Acknowledgment}
We thank the Defense Threat Reduction Agency (DTRA) for support.

\section*{References}

\begin{hangref}

\item Aiex, R.M. 2002. Uma investigacao
  experimental da distribuicao de probabilidade de tempo de
  solucao em heuristicas GRASP e sua aplicacao
  na analise de implementacoes paralelas.  PhD thesis,
  Department of Computer Science, Catholic University of Rio de
  Janeiro, Rio de Janeiro, Brazil.

\item Aiex, R.M., M.G.C. Resende, C.C. Ribeiro. 2002. 
  Probability distribution of solution time in GRASP: An experimental
  investigation. {\it J. Heuristics} {\bf 8} 343-373.

\item Anstreicher, K. 2003. Recent advances in the solution
  of quadratic assignment problems. {\it Math. Program.} {\bf 97}, 27-42.

\item Burkard R.E., E. Cela, S. E. Karish, F. Rendl. 1997.
  QAPLIB - A Quadratic Assignment Problem Library. {\it J.Global
  Optim.} {\bf 10} 391-403; http://www.seas.upenn.edu/qaplib/

\item Cela E. 1998. {\it The Quadratic Assignment Problem: Theory and
    Algorithms}. Kluwer, Boston, MA.

\item Helsgaun K. 2000. An Effective Implementation of the
  Lin-Kernighan Traveling Salesman Heuristic. {\it Eur. J.
    Oper. Res.} {\bf 126} 106-130.

\item James T., C. Rego, F. Glover. 2009. Multistart Tabu Search
  and Diversification Strategies for the Quadratic Assignment
  Problem. {\it IEEE Tran. on Systems, Man, and Cybernetics
  PART A: SYSTEMS AND HUMANS} {\bf 39} 579-596.

\item Koopmans T., M. Beckmann. 1957. Assignment problems
  and the location of economic activities. {\it Econometrica} {\bf 25}
   53-76.

 \item Lin S., B.W. Kernighan. 1973. An effective heuristic algorithm
   for the traveling-salesman problem. {\it Operations Research} {\bf 21}
   498-516; http://www.cs.bell-labs.com/who/bwk/tsp/index.html.

\item Misevicius, A. 2005. A tabu search algorithm for the
  quadratic assignment problem. {\it Comput. Optim. and
  Appl.} {\bf 30} 95-111.

\item Misevicius, A. 2008 An implementation of the iterated tabu
  search algorithm for the quadratic assignment problem. Working
  Paper, Kaunas University of Technology, Kaunas, Lithuania.

\item Oliveira, C.A.S., P.M. Pardalos, G. Mauricio, C. Resende. 2004. GRASP
  with Path-Relinking for the Quadratic Assignment Problem.
  Ribeiro, C.C. and Martins, S.L. eds., {\it Efficient and Experimental
  Algorithms, LNCS} {\bf 3059}, 356-368, Springer-Verlag, Berlin.

\item Pardalos, P.M., F. Rendl, H. Wolkowicz. 1994. The quadratic
  assignment problem: A survey and recent developments.  Pardalos,
  P.M., Wolkowicz, H., eds.  {\it Quadratic Assignment and Related
    Problems. DIMACS Series on Discrete Mathematics and Theoretical
    Computer Science} {\bf 16}, Amer. Math. Soc., Baltimore, MD.  1-42.

\item Taillard, E.  1991.  Robust taboo search for the quadratic
  assignment problem. {\it Parallel Comput.} {\bf 17} 443-455..


\end{hangref}

\newpage

\begin{table}
  \caption{Numerical Results.  {\it Improvement threshold} is the approximate target value below which the hybrid solution provides a performance improvement $>1$. {\it Target value} is the target value for which performance improvement is measured. $^a$ Misevicius (2005); $^b$ Misevicius (2008);  $^c$ current work. }
  \begin{tabular}{| l | r | r | r | r |}
\hline
  Problem & Best known  & Improvement & Target &Performance  \\
          &  solution   &   threshold    &  value &improvement \\
\hline
  Tai60a & 7205962$^a$ & 7320000 &  7256000 & 1.30 \\

  Tai80a & 13511780$^b$  & 13720000 & 13620000 & 2.52 \\

  Tai100a & 21052466$^b$  & 21360000 & 21200000 & 3.07 \\

  Pau200a & 89282330$^c$  & 89740000 & 89460000 & 10.94 \\

  Pau400a & 366463098$^c$  & 367600000 & 367060000 & 15.15\\
\hline
\end{tabular}

\label{table1}
\end{table}

\newpage

\end{document}